# USER-CENTRIC OPTIMIZATION FOR CONSTRAINT WEB SERVICE COMPOSITION USING A FUZZY-GUIDED GENETIC ALGORITHM SYSTEM

Mahdi Bakhshi[1] and Dr. Seyyed Mohsen Hashemi[2]

[1]Department of Computer Engineering, Islamic Azad University,
Shahrbabak Branch
Shahrbabak, Iran
`mb@shahrbabakiau.ac.ir`
[2]Department of Computer Engineering, Islamic Azad University,
Science and Research Branch, Tehran, Iran
`hashemi@isrup.com`

*ABSTRACT*

*Service-Oriented Applications (SOA) are being regarded as the main pragmatic solution for distributed environments. In such systems, however each service responds the user request independently, it is essential to compose them for delivering a compound value-added service. Since, there may be a number of compositions to create the requested service, it is important to find one which its properties are close to user's desires and meet some non-functional constraints and optimize criteria such as overall cost or response time. In this paper, a user-centric approach is presented for evaluating the service compositions which attempts to obtain the user desires. This approach uses fuzzy logic in order to inference based on quality criteria ranked by user and Genetic Algorithms to optimize the QoS-aware composition problem. Results show that the Fuzzy-based Genetic algorithm system enables user to participate in the process of web service composition easier and more efficient.*

## 1. INTRODUCTION

Service composition is a main problem in service based environment. Service composition means how the simple services aggregate to construct a new compound service with more value. During several years ago, many researchers have worked on this problem. Heretofore, the diverse techniques have been presented based on different aspects for performing service composition [2],[3],[4],[5]. From a business view, it is so important to find a composition whose cost is lower than all other feasible compositions can be made up. In this paper we are going to find an approach in order to select the optimal composition among different feasible compositions, according to quality criteria of services by creation a Fuzzy-guided Genetic Algorithm System (FGS).

One composite service performs specific functionalities which can be divided into some component functions. Also, they can be accomplished by some component services respectively. An example of a composite service is shown in Fig. 3 of [6]. The relations between component

DOI : 10.5121/ijwsc.2012.3301             1

functions are represented with the state chart in [3]. Some candidate services with same functionality and different QoS values (non-functionality) are discovered for each task (abstract service). Thus, there are various execution plans for each execution path in order to execute one composite service. Furthermore, since the number of candidate services with same functionality and different QoS values is increasing with the proliferation of web services, the composite size should be larger and larger. For example, in one execution path, there are 10 component function (task) and 20 candidate web services for each component function. In this composition scenario, the size of composite service should be about $20^{10}$. Since users who request web services usually have both functional requirements and global QoS constraints, it is necessary to select candidate services for a given task to achieve the best composite service and maximize user satisfaction (here, the expression "the best" means the composite service which has the optimal QoS values). Thus, web service selection with global QoS constraint satisfaction performs an important role in the process of web service composition [7],[8].

There are several approaches for QoS-aware web service composition. But most approaches are concerned about web service composition algorithm itself, while ignoring the flexibility for user to set QoS and cost. Most of them require QoS constraints given in form of numbers. In reality, it is difficult for users because they don't know the exact value or range of QoS of composite web services [25]. In this paper we try solve this problem through getting user constraints in form of fuzzy constraints.

The remainder of this paper is organized as follows. After a review of performed works in section 2, we review the literature of QoS-aware web service composition using QoS computation in Section 3. Section 4 presents modeling user preferences using quality driven fuzzy rules, designs a Fuzzy-guided Genetic Algorithm System (FGS) in order to select the optimal web service composition according to user preferences. Section 5 reports and discusses the results obtained from the simulations. Finally, Section 6 concludes.

## 2. RELATED WORK

So far, different approaches are used for selecting the optimal composition of services from quality properties point of view, like simple additive weighting technique (SAW) that stated in [14]. But, the stated work in [3] is an effort based on using of linear programming (LP) technique based on using constraints influence and introducing objective function for compositions measurement. A survey of some nonlinear approaches is discussed in [15]. Also, genetic algorithms are proposed for modeling composite services in the form of population members without any supposition on linearity of problem contents [13].

The computation of QoS values based on QoS matrix is an appropriate solution. Web services were ranked by normalizing QoS matrix in [16]. Anyway, it was only a local optimization algorithm but not a global one for service selection problem. Other works in the area of QoS computation include [3],[16], which proposed local optimization and global planning. The local optimization technique could not take global QoS constraints into consideration. When the size of composite service is very large, for example $20^{10}$, the overhead of global planning is quite enormous. Hereby, both had limitation to some extent. The mentioned techniques aren't able to effectively solve the web service selection issue with global QoS constraints. This kind of issues is NP-hard [16].

GA is a more suitable way in order to solve such issues. But GA performs an important role when the size of composition is very large. In [13], some numerical simulations show that linear Programming outperforms GA when the combinatorial size is small. Thus, GA should be preferred instead of linear Programming in the case of widely used services. On the other hand, linear Programming is to be preferred in the case of very specific services.

Generally, there are a few approaches that have investigated how various attributes of workflows or composite web services can be aggregated [12],[17]. The proposed approach in this paper performs selecting of optimal composition by added intuition of these works. Also, in this paper we have shown, that ranking and thus plan selection is a possible use case for aggregation of web service attributes, and we have also shown that how ranking can be calculated and plan selection can be automated.

Therefore, there are some techniques for selecting the optimal service from the point view of the user, based on giving high score to the service such as, whatever is proposed in [18] that includes some approaches for generating suitable fuzzy rules for the service selection problem. But, whatever is proposed in [19] tries to model user preferences by fuzzy rules in different strategies in order to selecting the optimal service according to user preferences. Most of the existing approaches for automatic selection of services, either consider only atomic services or they are not based on user preferences.

For independent global constraint web services composition problem, [26] presents an optimization method of web service composition with constraints using fuzzy Petri net (FPN), which can transform solving the optimal service composition problem into locating the largest trust value of legal firing sequences in the FPN model. Also To solve non-clarity and diversity of user's QoS requirements, a multi-strategic approach of fast composition of web services [25] is proposed. Finally, in [27] an improved version of the standard genetic algorithm approach by using fuzzy logic during the stochastic genetic search process is proposed. The fuzzy component dynamically adjusts the crossover and mutation evolution rates for each ten consecutive generations.

## 3. COMPUTING THE QOS OF COMPOSITE SERVICES

Services are constitutive units of service oriented systems. The service is presented serve to the service receptor by server and can be recalled by service receptor. These services can be combined and produce a value-added service. A composite service is an umbrella structure aggregating multiple other elementary and composite web services, which interact with each other according to a process model [3]. The composition of services can lead to a predetermined objective, while don't become certain by elementary services.

According to Std. ISO 8402 [20] and ITU E.800 [21], QoS may include a number of non-functional properties such as cost, response time, availability and reliability. Thus, QoS value of a composite service can be computed by fair computation of QoS of each composite service.
Methods for compute of quality criteria values are different. These values are used for computation of QoS of composite services. Here, there are some aggregation functions that are used to compute the QoS value of each composite service. Table 1 provides some of these aggregation functions for the execution plan $p$ [13].

Table 1. Aggregation functions for computation of quality values.

| QoSAttr. | Sequence | Switch | Flow | Loop |
|---|---|---|---|---|
| Time (T) | $\sum_{i=1}^{n} T(t_i)$ | $\sum_{i=1}^{n} p_i * T(t_i)$ | $Max\{T(t_i)_{i\in\{1...p\}}\}$ | $k * T(t)$ |
| Cost (C) | $\sum_{i=1}^{n} C(t_i)$ | $\sum_{i=1}^{n} p_i * C(t_i)$ | $\sum_{i=1}^{p} C(t_i)$ | $k * C(t)$ |
| Availability (A) | $\prod_{i=1}^{n} A(t_i)$ | $\sum_{i=1}^{n} p_i * A(t_i)$ | $\prod_{i=1}^{p} A(t_i)$ | $A(t)^k$ |
| Reliability (R) | $\prod_{i=1}^{n} R(t_i)$ | $\sum_{i=1}^{n} p_i * R(t_i)$ | $\prod_{i=1}^{p} R(t_i)$ | $R(t)^k$ |
| Custom Attr. (F) | $f_s(F(t_i))$, $i \in \{1...m\}$ | $f_B((p_i, F(t_i)))$, $i \in \{1...n\}$ | $f_F(F(t_i))$, $i \in \{1...p\}$ | $f_L(k, F(t))$ |

Namely, for a Sequence construct of tasks $\{t_1,..., t_m\}$, the Time and Cost functions are additive while Availability and Reliability are multiplicative. The Switch construct of Cases *1,…,n*, with probabilities $p_1,...,p_n$ such that $\sum_{i=1}^{i=n} p_i = 1$, and tasks $\{t_1,..., t_n\}$ respectively, is always evaluated as a sum of the attribute value of each task, times the probability of the Case to which it belongs. The aggregation functions for the Flow construct are essentially the same as those for the Sequence construct, except for the Time attribute where this is the maximum time of the parallel tasks $\{t_1,...,t_p\}$[11]. Finally, a Loop construct with *k* iterations of task *t* is equivalent to a Sequence construct of *k* copies of *t* [13]. Of course, this table includes a lot of quality attributes. As mentioned in the last line, other features are definable by user.

## 4. A FUZZY-GUIDED GA APPROACH ACCORDING TO USER PREFERENCES

By moving toward the age of information, a hypothesis can formulate the human knowledge in the systematic form, and introduce an approximate description that is reliable and analyzable. This important subject is applicable by a fuzzy system [1].

User's need to use considered services with different quality properties cause to user have a determinative role in the process of service composition. For example, the cost criterion may be the first grade importance for a user, but his need can be provide with a medium response time, and for other users these preferences are vice versa. The main problem is providing an approach for selecting the optimal composition of services according to user preferences and quality criteria of services and the aggregated values for each quality criterion. Our work is an approach that relies on the concept of domain ontology for description of services by specifying valid vocabulary and adding semantic concepts for description of services. These vague semantic descriptions located in the form of fuzzy rules and create a criterion to measurement of composite services, and then determine and measure the importance of each rule according to user's clear point of views. In fact, we provide an approach for giving score to composite services by entering the user's point of views in the process of fuzzy inference.

### 4.1. Definition of Variables and Membership Functions of the System

In many application domains, the transition between the memberships of an individual from one set to another is smooth. Consider, For example, height of a human. Small children grow, but when do they stop to be small? Such kinds of knowledge can be encoded using techniques such as fuzzy logic [19].

Vague knowledge, i.e. rules based on fuzzy logic, are also important from the aspect of evaluating values of attributes that have very complex dependencies with other attribute values [19]. The vague membership functions can be modeled in the form of some fuzzy sets. On the other hand, in simplest form, a domain ontology would specifies the valid vocabulary of describing (naming) functional and nonfunctional properties that are allowed to occur in service descriptions, but we need a domain ontology that be able to define categories of linguistic variables. For example, the response time could be described with the terms fast, normal, slow, very slow [18].

After the complete knowledge about linguistic variables, we can define the membership functions. In our work, we use the triangular and trapezoidal shapes for defining membership functions. According to expressed quality criteria, we define linguistic variables in the form of fuzzy sets based on domain ontology to describe web services, as defined in Figure 1. The reason of using triangular shapes for defining input variables and defining variable terms as a symmetrical form is permanent change at input membership functions and the distinction between different quality vectors.

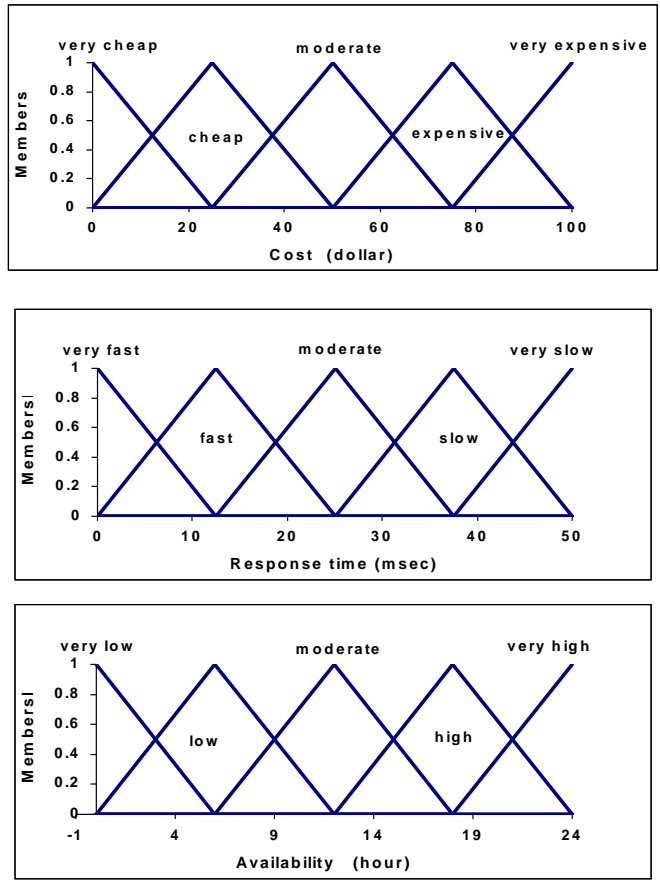

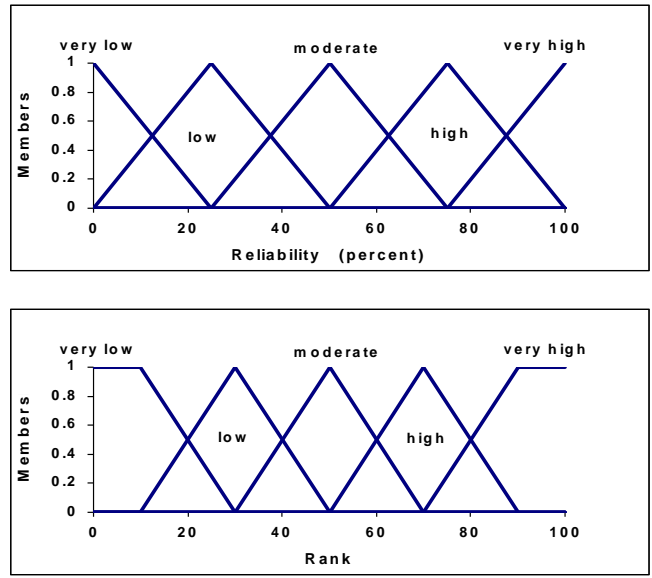

Figure 1. Membership functions for defining linguistic variables of the system.

In order to define membership functions, we use equal terms in definition of system's linguistic variables. The importance of this issue is because of logical relationship between the input and output variable terms in the formation of system's fuzzy rules.

### 4.2. Modeling User Preferences Based on Weighting the Rules

We regard preferences as the information that describes the constraints on the properties of an individual in order to be acceptable for further consideration. We specify different levels of acceptance by definition of fuzzy membership functions [19].

We model user preferences with fuzzy IF-THEN rules. Fuzzy IF-THEN rules allow to evaluate good approximations of desired QoS values in a very effective way [22],[23]. The IF part consists of membership function of various properties of an individual, and the THEN part is one of the membership functions of a special concept called Rank. Intuitively, a fuzzy rule describes which composition of attribute values a user is willing to accept to which degree, where attribute values and degree of acceptance are fuzzy sets, i.e. vague. An example of fuzzy rule can be:

*IF   Cost = Cheep   and   Response Time = Fast   THEN*
*Rank  =  High*

If all inputs classify into fuzzy sets viz. Very Low, Low, Moderate, High and Very High and The output Rank classifies as Very High, High, Moderate, Low and Very Low, then all possible combinations ($5^5$ i.e. 3125) of inputs are considered to design the rule base. Each rule corresponds to one of the five outputs based on the expert opinions. But modeling all of the preferences by users on fuzzy rules is very time consuming and maybe impossible.

Our approach with assumption existence of fuzzy rules that can be criteria for ranking quality vectors related to feasible execution plans, gives more weight to the rules that are more important from user's point of view.

The confidence factor (CF) of each rule which is a number between 0 and 1, can express the importance of a rule to obtain the final result. Equation (1) expresses the effect of this factor in the process of computing the result [24].

$$Membership_{con,i} = Membership_{premise,i} \times CF_i \tag{1}$$

This equation shows that the membership function of conclusion part in each rule $i$, is the result of multiplying membership function of premise part by confidence factor of the relevant rule.

We can provide the preliminary of fuzzy system with complete knowledge about the quality criteria and definition of input and output linguistic variables with equal terms. After that, we obtain some category of fuzzy rules for each quality criterion, which in each category there are some fuzzy rules that their number is equal to the number of input variable terms. In order to express fuzzy rules, we create one logical mapping between input variable terms in premise part and output variable in conclusion part for each category of rules. The effect of each rule in the ranking process should be distinct by user. This work is done by means of getting the importance grade of each quality criterion and located it as a confidence factor related to the one category of rules. Therefore we define the importance grade as a number from 0 to 100 and by conversion of distance is used as confidence factor.

In order to introduce fuzzy rules, we must create a logical mapping according to this point that whether low or high value of variable is considerable for user. The fuzzy rules for cost variable that low value of this variable is considerable for user can be expressed as follow:

$CF_{cost}$   IF   *Cost=very cheap*   THEN   *Rank=very high*
$CF_{cost}$   IF   *Cost=cheap*   THEN   *Rank=high*
$CF_{cost}$   IF   *Cost=moderate*   THEN   *Rank=moderate*
$CF_{cost}$   IF   *Cost=expensive*   THEN   *Rank=low*
$CF_{cost}$   IF   *Cost=very expensive*   THEN   *Rank=very low*

While, we express the fuzzy rules for availability variable that high value of this variable is considerable for user as follow:

*CFav*   IF   *Availability=very high*   THEN   *Rank=very high*
*CFav*   IF   *Availability=high*   THEN   *Rank=high*
*CFav*   IF   *Availability=moderate*   THEN   *Rank=moderate*
*CFav*   IF   *Availability=low*   THEN   *Rank=low*
*CFav*   IF   *Availability=very low*   THEN   *Rank=very low*

As pointed out above, we express (N=20) fuzzy rules for the fuzzy system that are according to the numbers of system linguistic variables and variable terms, the number of these rules are variable. These rules are criterion for evaluating different composite services.

**4.3. A Genetic Algorithm based Optimization**

By applying a GA-based approach the optimal solution (represented by its genotype)is determined by simulating the evolution of an initial population (through generation)until survival of best fitted individuals (here compositions) satisfying some constraints. The survivors are obtained by crossover, mutation, selection of compositions from previous generations. Details of GA parameterization follow:

- **Genotype:** it is defined by an array of integer. The number of items is equal to the number tasks involved in the composition. Each item, in turn, contains an index to an array of candidate services matching that task. Each composition, as a potential solution of the optimization problem, can be encoded using this genotype (e.g., Figure 2 is encoding the genotype of one composition).

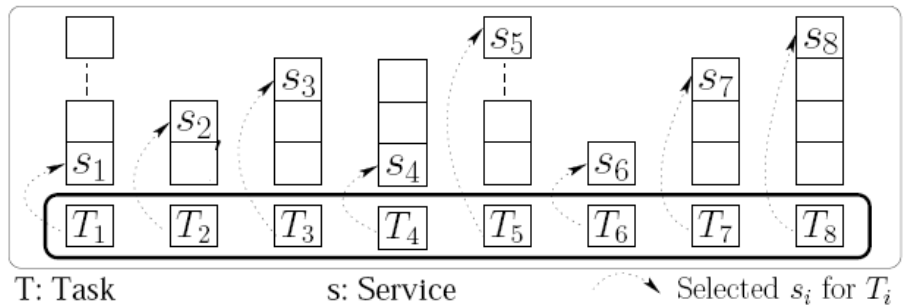

Figure 2. Genotype encoding for service composition.

- **Enhanced Initial Population:** The value of every task in every chromosome is set according to a local optimized method. The value of every task is QoS value of selected candidate service. The larger QoS value of a candidate service is, the larger the probability to be selected is. The probability of one candidate to be selected is the result of its QoS value divided by the sum of QoS values of all candidates of same task.

- **Fuzzy Constraints** have to be met by compositions c e.g., Cost have to be cheap, whereas the domain ontology it means Qco(c)<50 and Qco(c)>0 or response time at least have to be moderate that it means Qrt(c)<38 and Qrt(c)>10.

- **Fitness Function:** Now, the problem can be modeled by means of a fitness function and eventually, some constraints. The fitness function needs to maximize some QoS attributes (e.g., reliability), while minimizing others (e.g., cost). In our approach we define a new fitness function $f$ based defined fuzzy system.

$$f(c) = fit(input, Rank) \qquad (2)$$

Where input is quality vector of a composition c and Rank is output result of fuzzy system. $w_l \in [0,1]$ is the weight assigned to the $l^{th}$ quality criterion and $\sum_{l \in \{co, rt, av, re\}} w_l$ contrary to [10] doesn't have to be 1. Also QoS attribute factors ($Q_{l \in \{co, rt, av, re\}}$ don't have to be normalized in the interval [0,1]).

In addition $f$ must drive the evolution towards constraint satisfaction. To this end compositions that do not meet the constraints are penalized by (3).

$$f'(c) = f(c) - w_{pe} \sum_{l \in \{co, rt, re, av\}} (\frac{\Delta Q_l}{Q_l^{\max}(c) - Q_l^{\min}(c)})^2 \qquad (3)$$

Where $Q_l^{\max}$, $Q_l^{\min}$ are respectively the maximum and minimal value of the $l^{th}$ quality constraint, $w_{pe}$ weights the penalty factor and $\Delta Q_{l \in \{co, rt, re, av\}}$ is defined by:

$$\Delta Q_l = \begin{cases} Q_l - Q_l^{\max} & if \quad Q_l < Q_l^{\max} \\ 0 & if \quad Q_l^{\min} < Q_l < Q_l^{\max} \\ Q_l^{\min} - Q_l & if \quad Q_l^{\min} < Q_l \end{cases} \qquad (4)$$

Contrary to [5], compositions that violate constraints do not receive the same penalty. Indeed the factor $w_{pe}$ is further penalized in (3). This function avoids local optimal by considering also compositions that disobey constraints. Unfortunately, (3) contains a penalty for candidate compositions, which is the same at each generation. If, as usual, the weight $w_{pe}$ for this penalty factor is high, there is a risk that also candidate composition violating the constraints but "close" to a good solution could be discarded.

The alternative is to adopt a dynamic penalty, i.e., a penalty having a weight that increases with the number of generations. This allows, for the early generations, to also consider some individuals violating the constraints. After a number of generations, the population should be able to meet the constraints, and the evolution will try to improve only the rest of the fitness.
gen is the current generation, while maxgen is the maximum number of generations.

• **Operators on Genotypes:** they define authorized alterations on genotypes not only to ensure evolution of compositions' population along generations but also to prevent convergence to local optimum. We use: i) composition mutation i.e., random selection of a task (i.e., a position in the genotype) in a candidate composition and replacing its service with another one among those available, ii) the standard two-points crossover i.e., randomly combination of two compositions and iii) selection of compositions which is fitness-based i.e., compositions disobeying the constraints are selected proportionally from previous generations.

• **Stopping Criterion:** it enables to stop the evolution of a population. First of all we iterate until the constraints are met (i.e., $\Delta Q_{l=0} \; \forall l \in \{co, rt, re, av\}$) within a maximum number of generations. Once the latter constraints are satisfied we iterate until the best fitness composition remains unchanged for a given number of generations.

### 4.4. Design of System

Figure 3 shows an aspect of the system. The system has several components which are described below. The complete knowledge about quality criteria and then defining linguistic variables and membership functions has a determinative role in fuzzification process of composite services. The set of fuzzy rules which are proposed based on the logical mapping between the terms of linguistic variables, are the criteria for evaluating the different composite services. But, these rules are completely neutral against previous approaches of selecting suitable composition. Therefore, the user's role for preferring the rules that express his needs increases. As observed in the figure, the received user preferences are based on the importance grade that is given to each quality criterion. Then by changing distance, this numbers stated as confidence factors or weight of each category of rules.

The plan generation unit produces all feasible plans based on workflow and presents services for doing tasks. These plans can be limited by user constraints. For example, at composite service of travel planner user can determine the maximum cost that he can pay for hotel or car rent and so, infeasible execution plans will be omitted. In our system (FGS) these end compositions that do not meet the constraints are penalized, but because they may be close to a good solution, they are not discarded. On the other hand, aggregation functions for computing QoS of execution plans formed quality gathering unit which create quality vector of each execution plan.

Finally, there is optimization of composition unit which works based on GA. Fuzzy rules, which are created based on user preferences together with user constraints constitute one fitness function. GA parameters are defined for the system. Quality vectors related to different compositions are evaluated and one optimized solution is selected in accordance with user need and convenience.

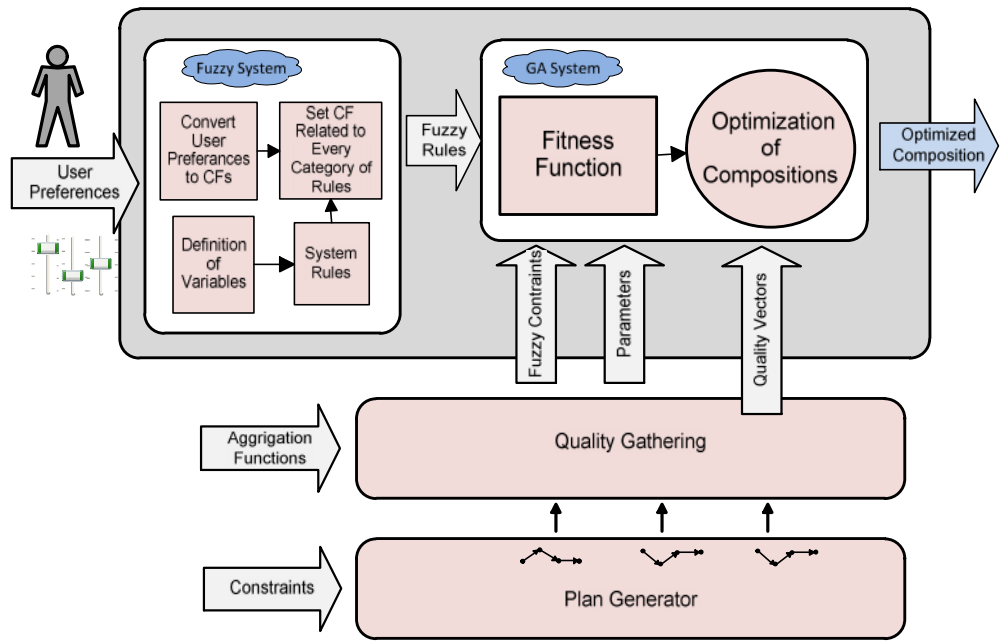

Figure 3. The general view of designed system.

## 5. EMPIRICAL STUDY

We implemented this approach by designing the FGS in MATLAB application and then run this at XP operating system. The optimal compositions are computed by using an elitist GA where the best 2 compositions were kept alive across generations, with a crossover probability of 0.7, a mutation probability of 0.1, a population of 200 compositions. The roulette wheel selection has been adopted as selection mechanism. We consider a simple stopping criterion i.e., up to 400

International Journal on Web Service Computing (IJWSC), Vol.3, No.3, September 2012generations. We conducted experiments on Intel(R) Core(TM)2 CPU, 2.4GHz with 2GB RAM.
We conducted experiments on Intel(R) Core(TM)2 CPU, 2.4GHz with 2GB RAM.

### 5.1. Evaluation of Fitness Function

At first, we set confidence factor of each quality criterion equal to 1 and draw charts related to changes of system variables. Figure 4 shows the changes of cost variable and results of these changes on rank of composite services. The output level from the changes of input variables, show the logical changes on rank of composite services.

Then, we decrease confidence factor that is related to this criterion (cost) and fix other criteria. Now, we can observe that the chart gradient and width of ranking scores in each step of confidence factor's reduction, is lessened. This subject is true for other criteria and is a reason for correct functionality of system. Figure 5 shows difference of maximum and minimum ranking scores belong to composite services against changes of confidence factor related to one quality criterion and fixing the other criteria in 1.

As observed here, the user's point of view has direct effect on computed score for a composite service. In fact, user can select a suitable composition through his point of view.

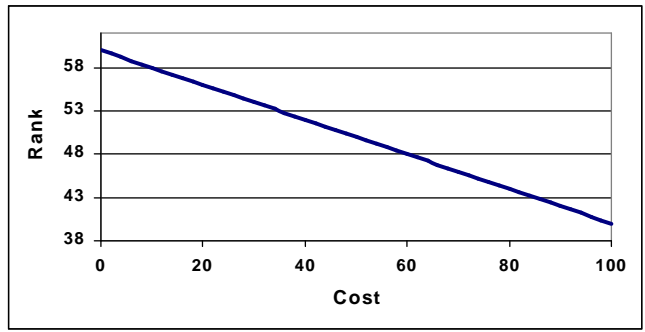

Figure 4. Effect of cost variable changing on rank variable.

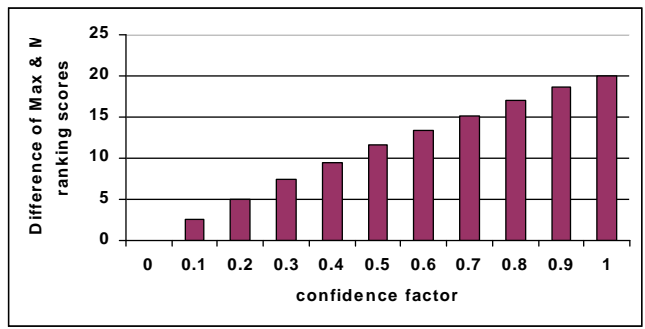

Figure 5. Effect of confidence factor changing on width of ranking scores.

### 5.2. Evolution of the Composition Quality

Figure 6 reports the evolution of the composition quality over the GA generations, by varying the number of tasks. For each task, we considered 30 available candidate services. This illustrates different levels of convergence to a composition that meets some constraints and optimizes its different quality criteria by maximizing the availability and reliability while keeping low cost and response time. For better evaluation, equal weights are assigned to the different quality criteria.

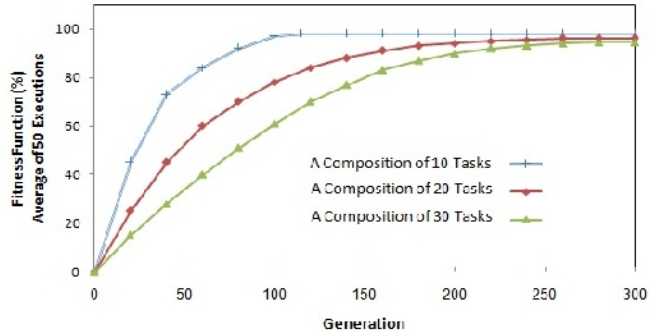

Figure 6. Evolution of the composition quality.

Table 2 and Figure 6 present the computation costs and the number of generations required to obtain the maximal fitness value. The more the number of tasks, the more the amount of time it takes to converge to the optimum. Obviously, the population size and the number of generations should be extended to reach the optimum of more complex compositions.

Table 2. Overview of computation costs.

| Tasks Num. | Max. Fitness (%) | Generation Num. | Time (ms) |
|---|---|---|---|
| 10 | 99 | 110 | 1230 |
| 20 | 97.5 | 267 | 2560 |
| 30 | 96 | 342 | 5540 |

### 5.3. Towards Large Scale Based Compositions

In this experiment we suggest to study the behavior of our approach regarding the optimization with a large number of tasks (up to 500 tasks) and candidate services (500). To this end we focus on its scalability and the impact of the number of generations as well as the population size on the GA success.

Table 3. Large scale compositions.

| Tasks Num. | Max. Fitness (%) | Generation Num./ Population size | Time (ms) |
|---|---|---|---|
| 100 | 87 | 400/200 | 4350 |
| | 98 | 700/400 | 9543 |
| 300 | 52 | 400/200 | 5836 |
| | 96 | 1500/500 | 20568 |
| 500 | 27 | 400/200 | 8056 |
| | 95 | 3000/1000 | 55655 |

As illustrated in Table 3, increasing both the number of generations and the population size does actually result in better fitness values for problems with a larger number of tasks and candidate services. For example, regarding the optimization of a composition of 500 tasks with 500 candidate services, a number of generations of 400 and a population size of 200 do result in a low fitness value of 27% of the maximum, whereas considering a number of generations of 3000 and a population size of 1000 achieve 95% of the maximum. Note that better fitness values can be reached by further increasing the sizes of generations and populations.

### 5.4. Convergence of GA-based approaches

In this experiment, we compare the convergence of FGS with the main alternative at present [10]. Also for each task, there is 30 available candidate services.

Table 4. Comparing GA-based approaches (population size of 200).

| Tasks Num. | Approach | Max. Fitness (%) | Generation Num. | Time (ms) |
|---|---|---|---|---|
| 10 | FGS | 99 | 110 | 1230 |
|  | [10] | 98 | 156 | 1350 |
| 20 | FGS | 97.5 | 267 | 2560 |
|  | [10] | 93 | 425 | 2865 |
| 30 | FGS | 96 | 342 | 5540 |
|  | [10] | 84 | 596 | 6570 |

According to Table 4, the advantage of FGS is twofold. Firstly we obtain better fitness values for the optimal composition than the approach of [10]. Secondly, our approach converges faster than the approach of [10]. In addition FGS avoids getting trapped by local optimums by i) further penalizing compositions that disobey constraints (the factor of $w_{pc}$ in (3) and (5)) and ii) suggesting a dynamic penalty, i.e., a penalty having a weight that increases with the number of generations. These results support the adoption of FGS in the cases where a large number of tasks and services are considered.

A linear increase in time for increasing numbers of iterations for the GA which is proposed in [10] and FGS are shown in Figure 7. Due to the additional time needed for the fuzzy component, FGS shows the larger increase, but because of need to less iteration it is more valuable.

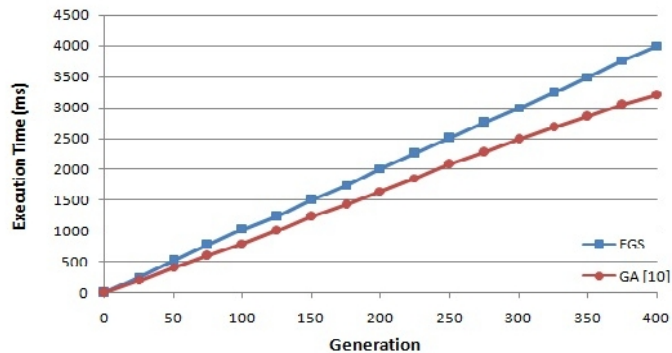

Figure 7. Execution times obtained using GA and FGS.

## 6. CONCLUSIONS

This paper proposed a hybrid Fuzzy-guided Genetic Algorithm system for QoS-aware service composition, i.e., to determine a set of candidate services to be bound to abstract services contained in a composition to meet a set of fuzzy constraints and to optimize a fitness criterion on QoS attributes. In the GA optimization, the fitness function is a fuzzy system that is constructed based on user preferences.

Several advantages can be stated for this approach. This approach emphasizes on accordance to the user preferences and quality properties of composite service. The user clearly states his preferences on the other hand he states many fuzzy rules. Also, in addition to high care in expression of preferences, for modeling different user preferences there is no need to restate the rules. User presents fuzzy constraints and don't have to know details of information about QoS values. Also, this system is extensible against increasing the quality criteria. Compared with GA, this approach has better fitness values and faster convergence and more accuracy.

Finally, in future work which is user-centric in order to select the optimal web service composition, we will consider composition of one technique such as simulated annealing or migrating birds with fuzzy logic and survey their results through further experiments.

## Authors

**M. Bakhshi** received his B.Sc in computer engineering from Shahid Bahonar University of Kerman, Iran, his M.Sc degree in software engineering from Islamic Azad University of Najaf Abad, Iran. Currently he is faculty of Islamic Azad University, shahrbabak branch. His interests include Web Service technology and coordination problem. He is working on dynamic choreography models for Web services in B2B Corporation.

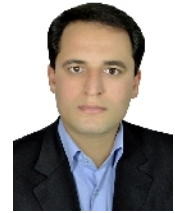

Dr. S. M. Hashemi received his M.S. degree in Computer Science from Amirkabir University of Technology in 2003, and his PhD degree in Computer Science from the Azad University in 2009. Moreover, he is currently a faculty member at Science and Research Branch, Azad University, Tehran. His current research interests include Software Intensive Systems, Global Village Services, Grid Computing, Agile Enterprise Architecting through ISRUP, and Globalization Governance through IT/IS Services.

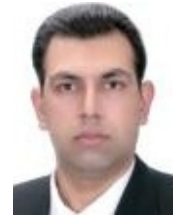